
\def\m@th{\mathsurround=0pt}
\newif\ifdtpt
\def\displ@y{\openup1\jot\m@th
    \everycr{\noalign{\ifdtpt\dt@pfalse
    \vskip-\lineskiplimit \vskip\normallineskiplimit
    \else \penalty\interdisplaylinepenalty \fi}}}
\def\eqalignc#1{\,\vcenter{\openup1\jot\m@th
                \ialign{\strut\hfil$\displaystyle{##}$\hfil&
                              \hfil$\displaystyle{{}##}$\hfil&
                              \hfil$\displaystyle{{}##}$\hfil&
                              \hfil$\displaystyle{{}##}$\hfil&
                              \hfil$\displaystyle{{}##}$\hfil\crcr#1\crcr}}\,}
\def\eqalignnoc#1{\displ@y\tabskip\centering \halign to \displaywidth{
                  \hfil$\displaystyle{##}$\hfil\tabskip=0pt &
                  \hfil$\displaystyle{{}##}$\hfil\tabskip=0pt &
                  \hfil$\displaystyle{{}##}$\hfil\tabskip=0pt &
                  \hfil$\displaystyle{{}##}$\hfil\tabskip=0pt &
                  \hfil$\displaystyle{{}##}$\hfil\tabskip\centering &
                  \llap{$##$}\tabskip=0pt \crcr#1\crcr}}
\def\leqalignnoc#1{\displ@y\tabskip\centering \halign to \displaywidth{
                  \hfil$\displaystyle{##}$\hfil\tabskip=0pt &
                  \hfil$\displaystyle{{}##}$\hfil\tabskip=0pt &
                  \hfil$\displaystyle{{}##}$\hfil\tabskip=0pt &
                  \hfil$\displaystyle{{}##}$\hfil\tabskip=0pt &
                  \hfil$\displaystyle{{}##}$\hfil\tabskip\centering &
                  \kern-\displaywidth\rlap{$##$}\tabskip=\displaywidth
                  \crcr#1\crcr}}

\def\charlvmidlw#1#2{\,\vtop{\ialign{##\crcr
      #1\crcr\noalign{\kern1pt\nointerlineskip}
      $\hfil#2\hfil$\crcr}}\,}
\def\charlvlowlw#1#2{\,\vtop{\ialign{##\crcr
      $\hfil#1\hfil$\crcr\noalign{\kern1pt\nointerlineskip}
      #2\crcr}}\,}
\def\charlvmidup#1#2{\,\vbox{\ialign{##\crcr
      $\hfil#1\hfil$\crcr\noalign{\kern1pt\nointerlineskip}
      #2\crcr}}\,}
\def\charlvupup#1#2{\,\vbox{\ialign{##\crcr
      #1\crcr\noalign{\kern1pt\nointerlineskip}
      $\hfil#2\hfil$\crcr}}\,}

\def\vspce{\kern4pt} \def\hspce{\kern4pt}

\def\emptybox{\vbox{\kern.7ex\hbox{\kern.5em}\kern.7ex}}
 \font\sevmi  = cmmi7
    \skewchar\sevmi ='177
 \font\fivmi  = cmmi5
    \skewchar\fivmi ='177
\font\tenmib=cmmib10
\newfam\bfmitfam

\textfont\bfmitfam=\tenmib
\scriptfont\bfmitfam=\sevmi
\scriptscriptfont\bfmitfam=\fivmi

\def\mathcedilla{\vtop{\hbox{c}{\kern0pt\nointerlineskip}
	         {\hbox{$\mkern-2mu \mathchar"0018\mkern-2mu$}}}}

\mathchardef\gq="0060
\mathchardef\dq="0027
\mathchardef\ssmath="19
\mathchardef\aemath="1A
\mathchardef\oemath="1B
\mathchardef\omath="1C
\mathchardef\AEmath="1D
\mathchardef\OEmath="1E
\mathchardef\Omath="1F
\mathchardef\imath="10
\mathchardef\fmath="0166
\mathchardef\gmath="0167
\mathchardef\vmath="0176

\def\twodot{.\kern-0.1em.}

\def\paral{\mathrel{/\kern-.25em/}}
\def\grlo{\mathrel{\hbox{\lower.2ex\hbox{\rlap{$>$}\raise1ex\hbox{$<$}}}}}
\def\logr{\mathrel{\hbox{\lower.2ex\hbox{\rlap{$<$}\raise1ex\hbox{$>$}}}}}
\def\greq{\mathrel{\hbox{\lower1ex\hbox{\rlap{$=$}\raise1.2ex\hbox{$>$}}}}}
\def\loeq{\mathrel{\hbox{\lower1ex\hbox{\rlap{$=$}\raise1.2ex\hbox{$<$}}}}}
\def\grsim{\mathrel{\hbox{\lower1ex\hbox{\rlap{$\sim$}\raise1ex\hbox{$>$}}}}}
\def\losim{\mathrel{\hbox{\lower1ex\hbox{\rlap{$\sim$}\raise1ex\hbox{$<$}}}}}
\font\ninerm=cmr9
\def\uniset{\rlap{\ninerm 1}\kern.15em 1}

\def\emptysq{\mathbin{\vbox{\hrule\hbox{\vrule height1ex \kern.5em
                            \vrule height1ex}\hrule}}}
\def\emptyrect{\mathbin{\vbox{\hrule\hbox{\vrule height1ex \kern1em
                              \vrule height1ex}\hrule}}}
\def\rightonleftarrow{\mathrel{\hbox{\raise.5ex\hbox{$\rightarrow$}\ignorespaces
                                   \lower.5ex\hbox{\llap{$\leftarrow$}}}}}
\def\leftonrightarrow{\mathrel{\hbox{\raise.5ex\hbox{$\leftarrow$}\ignorespaces
                                   \lower.5ex\hbox{\llap{$\rightarrow$}}}}}

\def\bkB{{\rm I\kern-.17em B}}
\def\bkC{{\rm \kern.24em
            \vrule width.05em height1.4ex depth-.05ex
            \kern-.26em C}}
\def\bkD{{\rm I\kern-.17em D}}
\def\bkE{{\rm I\kern-.17em E}}
\def\bkF{{\rm I\kern-.17em F}}
\def\bkG{{\rm \kern.24em
            \vrule width.05em height1.4ex depth-.05ex
            \kern-.26em G}}
\def\bkH{{\rm I\kern-.22em H}}
\def\bkI{{\rm I\kern-.22em I}}
\def\bkJ{{\rm \kern.19em
            \vrule width.02em height1.5ex depth0ex
            \kern-.20em J}}
\def\bkK{{\rm I\kern-.22em K}}
\def\bkL{{\rm I\kern-.17em L}}
\def\bkM{{\rm I\kern-.22em M}}
\def\bkN{{\rm I\kern-.20em N}}
\def\bkO{{\rm \kern.24em
            \vrule width.05em height1.4ex depth-.05ex
            \kern-.26em O}}
\def\bkP{{\rm I\kern-.17em P}}
\def\bkQ{{\rm \kern.24em
            \vrule width.05em height1.4ex depth-.05ex
            \kern-.26em Q}}
\def\bkR{{\rm I\kern-.17em R}}
\def\bkT{{\rm \kern.24em
            \vrule width.02em height1.5ex depth 0ex
            \kern-.27em T}}
\def\bkU{{\rm \kern.30em
            \vrule width.02em height1.47ex depth-.05ex
            \kern-.32em U}}
\def\bkZ{{\rm Z\kern-.32em Z}}

\magnification=1200
\pageno=1
\baselineskip=17pt
\hfill{SPhT T95/032}

\hfill{LPTHE -95/29}
\bigskip
\bigskip
\centerline{{\bf
INITIAL-STATE COLOUR DIPOLE EMISSION ASSOCIATED WITH}}

\centerline{{\bf QCD POMERON  EXCHANGE}}

\vglue 1truecm

\centerline {{\bf A. Bialas}
\footnote{$ ^{a)}  $}{
CEA, Service de Physique Th\'eorique, CE-Saclay
F-91191 Gif-sur-Yvette Cedex, FRANCE}
\footnote{$ ^{b)      }$}{
LPTHE, Universit\'e Paris-Sud,
91405 Orsay Cedex, FRANCE
Laboratoire associ\' e au CNRS, URA-D0063}
\footnote{$ ^{\dag} $}{On leave from Theoretical
Physics Department, Jagellonian
University, 30059 Krakow, Reymonta 4, POLAND}
and {\bf R. Peschanski} {$ ^{a)}  $}}

\vglue 2truecm

\centerline{ABSTRACT}

The initial-state radiation of soft colour dipoles produced together with
a single QCD
Pomeron exchange (BFKL) in onium-onium scattering is calculated in the
framework of  Mueller's approach. The resulting dipole production grows
with increasing energy and reveals an unexpected feature of a power-law tail
at appreciably large transverse distances from the collision axis, this
phenomenon being
related to the scale-invariant structure of dipole-dipole correlations.

\vfill\eject

High energy onium-onium scattering is a simple process which can be used to
study the physics of the perturbative QCD Pomeron, the so-called BFKL
singularity$ ^{\lbrack 1\rbrack} . $ Recently a quantitative picture
in which a high-$ Q^2 $ $ q\bar q $ (or {\it onium}) state
looks like a
collection of colour dipoles of various sizes
has been developped
by  Mueller$ ^{\lbrack 2-4\rbrack} $
. The QCD Pomeron
elastic amplitude is recovered in this dipole picture provided the onium-onium
elastic scattering comes from a dipole in one onium state scattering off a
dipole in the other onium state by means of two-gluon exchange$ ^{\lbrack
2-4,5\rbrack} . $In this paper we elaborate some consequences
of the dipole approach for the initial-state radiation associated with a
single Pomeron exchange.

Our starting point is the observation that the onium-onium
scattering process is accompanied by a radiation due to colour dipoles
which -- while present in the initial state -- are released during the
collision. In this note we present an explicit computation of
this dipole emission process
following  Mueller's approach, and more
specifically that of Ref.$\lbrack$4$\rbrack$.
We have found that the
cross-section for the production
of dipoles emitted from one of the colliding onia can be approximated by:
$$ { {\rm d} \sigma \over {\rm d} x {\rm d}^2r} = \int^{ }_{ } {\rm d}^2x_{10}
{\rm d}^2x^{\prime}_{ 10} \int^ 1_0 {\rm d} z\ {\rm d} z^{\prime} \ \phi
\left(x_{10},z \right) \phi \left(x^{\prime}_{ 10},z^{\prime} \right) { {\rm
d}\hat
\sigma \over {\rm d} x {\rm d}^2r}, $$
where
$$
{ {\rm d}\hat \sigma \over {\rm d} x
{\rm d} r^2}
 = \ \sigma_{ {\rm tot}}\
{1 \over xr^2}\ {\rm e}^{\left(
\left(\alpha_ p-1 \right){Y \over 2} \right)}\ {x_{10} \over x} \left({r \over
x_{10}} \right)^{\gamma_M} \varphi \left({x x_{10} \over r^2}
\right)\ , \eqno (1) $$
where$^{[4]}$ $\sigma_{{\rm tot}} = 2 \pi
\ x_{10}x^{\prime }_{10}\ \alpha ^2
\ {\rm e}^{(\alpha_p-1)\ Y}\ \left({2 a \over \pi}\right)^{1/2},$
  $ \alpha_ p
= 1+{\alpha N_c \over \pi} \ 4 {\rm ln}  2  $ is the intercept
of the QCD
Pomeron singularity (BFKL) and
$ \varphi$ is a function which acts as a cut-off on the scaling
behaviour $ \left({r \over x_{10}} \right)^{\gamma_M}. $
Within a high-energy approximation, this function can be
parametrized as follows:
$$ \varphi  \approx {\cal C} \
\left({2a \over \pi}
\right)^{3/2}
\ {\rm ln} \left({r^2 \over xx_{10}} \right) {\rm exp} \left({-a\
{\rm ln}^2 \left({r^2 \over xx_{10}} \right)}\right)
\ , \eqno (2) $$
where $ \cal C $ is a constant and the cut-off scale is defined as:
$$ a \equiv a(Y) = \left[7\alpha \ N_c\zeta( 3)Y/\pi \right]^{-1}\
\approx \ \ \left[ 3 \left(\alpha_ p-1 \right)Y\right]^{-1}
 \eqno (3) $$

In our result (1), and following the notations of Refs.$\lbrack$2-4$\rbrack$,
 $
\phi \left(x_{10},z \right) $ defines
the square of the heavy $ q\bar q $ component of the onium wave-function with
transverse-coordinate separation $ x_{10} $ and light-cone momentum $ z $ of
the
antiquark with respect to the onium; $ x $ is the transverse size of the
emitted
dipole, $ \vec r \ $ its transverse coordinate and $ e^{-Y/2} =
{P_{+}\over E_{c.m.}}$ denotes the
light-cone momentum fraction
of the softest gluon involved in constituting the emitted dipole.
As will be explained further on, the
important parameter $ \gamma_ M\approx .37 $ is a solution$
^{\lbrack 4\rbrack}$ of the equation
$$ \chi \left(\gamma_ M \right) \equiv  2\psi( 1) - \psi \left(1-\gamma_M/2
\right)-\psi \left(\gamma_M/2 \right) =
2\ \chi (1)
, \eqno (4) $$
where $ \psi( \gamma)  \equiv  {{\rm d}
\over {\rm d}
\gamma}
\ {\rm ln} \ \Gamma
,\ \chi (1)\ \equiv \  4 {\rm ln}  2
. $  The formula (1) is
expected to be valid provided
$$ r \gg  x,x_{10}\ ;\ \ \ \ 0 <
{\rm ln}  {r \over x},
{\rm ln}  {r^2 \over x_{10}x}\
< \ a^{- 1/2}(Y) \ ,
\eqno (5) $$
requiring large distances with respect to dipole sizes while limited by the
total amount of c.m.s. energy.

Before we proceed to outline the derivation of Eqs.(1-3), let us point out
their two interesting features.

\vskip 10pt

(a) for fixed $ x\ {\rm and} \ r $ the density of emitted dipoles increases as
a power of
the
 incident energy
$$  {1 \over \sigma_{ {\rm tot}}}
\ { {\rm d}\hat \sigma \over {\rm d} x {\rm d}^2r} \sim
\left({E_{c.m.}\over P_{+}}\right)^{
\alpha_
p-1} \equiv
\left({E_{c.m.}\over P_{+}}\right)^{
 \chi (1)     \  {\alpha N_c \over \pi}}
$$
with a power fully determined by the QCD Pomeron intercept $ \alpha_ p. $
Thus, the
contribution of soft-gluon radiation is increasing with energy and very
likely dominant at high enough energy at least in the region where the
QCD Pomeron exchange is relevant. Note that this contributions increases also
when $P_+$ decreases.

\vskip 10pt

(b) For fixed dipole size $ x $ the distribution reveals a {\it power-law}
tail
in the transverse position coordinate $ r, $ namely
$$ xr^2 { {\rm d} \sigma \over {\rm d} x {\rm d}^2r} \sim \ r^{\gamma_M}\ ,
\eqno (6) $$
which is expected to extend until a distance cut-off defined by Eqs.(2,3),
that is:
$$ r^2 \leq  r^2_{ {\rm max}} = x_{10}x\ {\rm e}^{a^{-1/2}} \approx  x_{10}x\
{\rm exp}\left(\sqrt{3(\alpha_p-1)Y}\right)
\eqno (7) $$
One sees that $ r_{ {\rm max}} $ increases as the exponential of
$\sqrt{{\rm ln} s}.$
 Thus -- at high enough energies -- the power-law behaviour (6) is
valid up to  distances appreciably exceeding the initial $ \left(x_{10} \right)
$ and emitted
($ x) $ dipole sizes. Since other contributions to gluon emission are
not
expected to possess this large distance tail, we infer that
the power-law tail (6) should be a dominant component of
gluon emission at relatively large distances from the collision axis.

${\bf 2.}$ Let us now outline the derivation of Eqs.(1-4). As
 already explained, we intend to estimate the emission of
colour dipoles from the initial-state. To
this end, we first write the formula for the inclusive cross-section
for emission of a dipole from one of the colliding onia. It reads:

$$ \eqalignno{x\
{ {\rm d}\hat \sigma \over {\rm d} x {\rm d} r^2}
& = 4\pi
\alpha^ 2 \int^{ \infty}_ 0{ {\rm d} l \over l^3}\ { {\rm d}\bar x \over\bar x}
\ { {\rm d}\bar x^{\prime} \over\bar x^{\prime}}  \left[1-J_0 \left(l\bar x
\right) \right] \left[1-J_0 \left(l\bar x^{\prime} \right) \right] &  \cr  &
\times  \int^{ }_{ } {\rm d}^2b\ {\rm d}^2b^{\prime} \ n^{(1)}
\left(x^{\prime}_{ 01},\vec b^{\prime} - \vec b,\bar x^{\prime} ;{Y\over
2}\right)n^{(2)}
\left(x_{01},\vec b,\bar x;\vec r,x;{Y\over 2} \right) & (8) \cr} $$
where $ n^{(1)} $ is the {\it single} dipole density in the onium of size $
x^{\prime}_{ 01} $ and $ n^{(2)} $
is the {\it double} dipole density in the onium of size $ x_{01}. $

This formula -which is an extension of Mueller's
formula for the total onium-onium cross-section (see Eq.(3) of Ref.[4])-
expresses simply the fact that the number of emitted dipoles
is just the number of dipoles present in the initial state
{\it whenever} the interaction took place. The process
is
illustrated
in Figure 1 where the geometry of the reaction in the plane transverse
to the collision axis is represented.
The important feature of Eq.(8) is that it contains the double-dipole density
in one of the initial onia (since one of the two dipoles
is involved in the interaction mechanism). Thus, it is sensitive to
{\it correlations} between dipoles in the same onium state. As
we shall see, this has non-trivial consequences. In fact, already
 at this stage, one may expect that these correlations should be
rather strong because, as shown in Refs.[2-4], the colour dipoles
which contribute to the onium wave function are formed in a cascade
process and thus cannot be independent. Furthermore, since this cascade is
scale-invariant one expects also scale invariance in the dipole-dipole
correlations which -in turn- is likely to be a reflection of the
inner conformal invariance, known to be rooted in the formalism
of the BFKL Pomeron$^{[8]}.$

The explicit expression for $ n^{(1)}, $ was given in Ref.[4]:
\vfill\eject
$$ \eqalignno{ n^{(1)} \left(x^{\prime}_{ 01},\vec b^{\prime} -\vec b,\bar
x^{\prime} ;Y/2 \right) & \equiv  {x^{\prime}_{ 01} \over 4x^{\prime}
\left\vert\vec b^{\prime} -\vec b \right\vert^ 2} {\rm ln}{ \left\vert\vec
b^{\prime} -\vec b \right\vert^ 2 \over x^{\prime}_{ 01} \bar x^{\prime}}
{\rm
exp}  \left(-a\ {\rm ln}^2{ \left\vert\vec b^{\prime} -\vec b \right\vert^ 2
\over
x^{\prime}_{ 01}\bar x^{\prime}}\right) \times  &  \cr  &  \times  {\rm exp}
\left[
\left(\alpha_{ p-1} \right)Y/2 \right] \left({4a \over \pi} \right)^{-3/2}\ ,
& (9) \cr} $$
where $ a = a(Y)$ is defined as in Eq.(3).

The explicit expression for $ n^{(2)} $ is not yet known, besides an
approximate
formula$ ^{\lbrack 4\rbrack} $ which is insufficient for our purpose of
deriving the overall
behaviour of $ { {\rm d}\hat \sigma \over {\rm d} x\ {\rm d}^2r} $ as a
function of the different variables of the problem.
To obtain an improved formula, valid at large distances from the colliding
onia, we employ an approximate form of the equation for $ n^{(2)}$
(given in [4]), with only terms dominant at large distances being
kept. Written for the Mellin-Transform $\tilde {n}$ of $n^{(2)}$
it reads:
$$ \eqalignno{
{ \partial\tilde n^{(2)} \over \partial Y}
\left(\gamma ,\vec
b,\bar x;\vec r,x;Y \right)
 & =
  {\alpha N_c \over \pi}
 \chi (\gamma)\ \tilde
n^{(2)} \left(\gamma ,\vec b;\vec r,x;Y \right)
 & (10) \cr  &
+ {\alpha N_c \over \pi^ 2}
\int^{ }_{ }
x^{1-\gamma }_{10} {\rm d} x_{10} \int^{ }_{ }
{ {\rm d}^2x_2 \over
x^2_{12}x^2_{02}}
 n^{(1)}
\left( x_{02},\vec
b,\bar x;Y \right)
n^{(1)} \left( x_{12},\vec r,x;Y \right)
 &  \cr} $$
where $ \chi (\gamma) $ is defined as in (4).
$  n^{(2)} $ may be recovered from the solution of (10)
by writing
$$ n^{(2)} \left(x_{01},\vec b,\bar x;\vec r,x;Y \right) = \int^{c+i\infty }_
{ c-i\infty}{
{\rm d} \gamma \over 2i\pi}  \ x_{10}^{\gamma}\ \tilde n^{(2)} \left(\gamma
,\vec b,\bar x;\vec
r,x;Y \right)\ . \eqno (11) $$
with the real constant $c < 2$ for convergence condition, and to the
right (in the complex $\gamma-$plane) of all singularities of $\tilde n^{(2)}.$

Let us first find an expression for the inhomogeneous term
 of Eq.(10)
denoted
$ \tilde n^{(2)}_0.
$
Using the analytic form (9) for the single-dipole distributions $ n_1 $ and
the
known Jacobian of the transformation of variables from $ \vec x_2 $ to $
x_{02}, $ $ x_{12}, $ namely$^{[2]}$
$$ {\rm d}^2x_2 \equiv  2\pi \ {\rm d} x_{02} {\rm d} x_{12}\ x_{02}x_{12}
\int^{ \infty}_ 0p {\rm d} p\ J_0 \left(px_{10} \right)J_0 \left(px_{12}
\right)J_0 \left(px_{02} \right)\ , \eqno (12) $$
one may write after some algebraic manipulations:
$$ \eqalignno{\tilde n^{(2)}_0
\left(\gamma ,\vec b,\bar x;\bar r,x;Y \right)
&
= {\rm exp} \left[2 \left(\alpha_ p-1 \right)Y
\right]\times {\cal G}
\left(\gamma ,\vec b,\bar x;\bar r,x;Y \right) & (13) \cr
{\cal G}
\left(\gamma ,\vec b,\bar x;\bar r,x;Y \right)
& = 2  {\alpha N_c \over \pi}   \left({2a \over \pi} \right)^3\ { 1 \over 16
\left(\bar xb^2
\right) \left(xr^2 \right)}  \int\int^{ }_{ } {\rm
d} x_{02} {\rm d} x_{12}\ {\rm ln} \left({b^2 \over x_{12}\bar x} \right)
{\rm ln} \left({r^2 \over x_{02}x} \right) &  \cr  &
\ \ \times \  {\rm exp}\left\{-{a \over 2} \left( {\rm
ln}^2{b^2 \over x_{12}\bar x}+ {\rm ln}^2{r^2 \over x_{02}x} \right)\right\}
W
\left(x_{02},x_{12} \right). &  \cr} $$
The function $ W \left(x_{01},x_{02} \right) $ can be expressed in terms of
generalized hypergeometric series
as follows:
$$ \eqalignno{
W \left(x_{02},x_{12} \right)
& \equiv  \int^{ \infty}_
0x^{1-\gamma}_{ 10} {\rm d} x_{10}\ p {\rm d} p\ J_0 \left(px_{10} \right)J_0
\left(px_{12} \right)J_0 \left(px_{02} \right) &  \cr  &  =
 \theta \left(x_{02}-x_{12} \right)\
x^{-\gamma}_{02}\ _2F_1 \left[ \left({x_{12} \over x_{02}} \right)^2
\right]
 + \{ 1 \longleftrightarrow  2\} \ ,
& (14) \cr} $$
where we used a shortened notation for the hypergeometric function$ ^{\lbrack
7\rbrack} $
$$ \ _2F_1 \left[y^2 \right] \equiv  \ _2F_1 \left({\gamma \over 2},
{\gamma \over
2};1;y^2 \right)\ . $$
{}From formula (13), it is quite clear that the dominant energy dependence of $
\tilde n_0 $ is given
by the exponential term $ {\rm exp} \left\{ 2 \left(\alpha_ p-1 \right)
\right\} . $ Consequently, we can approximately solve Eq.(10) for $ \tilde
n^{(2)} $
and obtain (in the limit $ Y \longrightarrow \infty ) $
$$ \tilde n^{(2)}(\gamma ,Y) \approx  {\rm e}^{{\alpha N_c \over
\pi} \chi(\gamma) Y}
\left(2 \left(\alpha_ p-1 \right)- {\alpha N_c \over \pi}  \chi( \gamma)
\right)^{-1} \times \ {\cal G}\ . \eqno (15) $$
Using (15) we can calculate the Mellin transform of
${ {\rm d}\hat \sigma \over {\rm d} x {\rm d} r^2},$ i.e.
$$
{{\rm d} \tilde  \sigma \over {\rm d} x {\rm d} r^2}\ (\gamma)
\ = \int_{ }^{ }\  \ {\rm d}x_{10}^{-\gamma-1}\
{ {\rm d}\hat \sigma \over {\rm d} x {\rm d} r^2}\ (x_{10})
\eqno (16)
$$

Indeed, inserting (8) and (13) into (16),
it is possible$^{[4]}$ to integrate over $ {\rm d}
l, $ $ {\rm d}\bar x, $ $ {\rm d}\bar x^{\prime} , $ $ {\rm d}^2b, $ $ {\rm
d}^2b^{\prime} $
since they decouple due
to     factorizability
properties of $ \tilde n^{(2)}_0. $ All in all, one gets
$$ \eqalignno{ { {\rm d}\tilde \sigma \over {\rm
d} x {\rm d} r^2} & = \hat \sigma_{ {\rm tot}} \times \ \sqrt{ 2}\ {\alpha N_c
\over \pi}\ {1 \over xr^2}\ {{\rm
exp} \left[
\left({\alpha N_c \over \pi} \chi (\gamma) - (\alpha_ p-1) \right)\ Y/2
\right] \over
2\ \left(\alpha_ p-1 \right)\
-\ {\alpha N_c \over \pi} \chi (\gamma)}
\left({2a \over \pi}
\right)^{3/2} & (17) \cr  &  \times
\int_0^r\int_0^r\  {\rm d} x_{12} {\rm d} x_{02}
\ {\rm ln}  {r^2 \over x_{12}x}\ {\rm e}^{-a {\rm ln}^2 \left({r^2
\over x_{12}x} \right)}\ W \left(x_{02},x_{12} \right)\ , &  \cr} $$
where$^{[4]}$ $\hat \sigma_{{\rm tot}} = {\sigma_{{\rm tot}}
\over x_{10}} = 2 \pi x^{\prime }_{10}\ \alpha ^2
\ {\rm e}^{(\alpha_p-1)\ Y}\ \left({2\ a \over \pi}\right)^{1/2}.$

With the explicit form of $
W \left(x_{02},x_{12} \right)
,$ Eq.(14),
the final integration over $ x_{02} $ and $ x_{12} $
can be performed.
Using the known asymptotic expansion of the error function$^{[7]}$ within
the approximations defined in (5),
one obtains as a final result:
$$ \eqalignno{ { {\rm d}\tilde \sigma \over {\rm
d} x {\rm d} r^2} & = \hat \sigma_{ {\rm tot}} (x^{\prime}_{10})\
{\cal H}(\gamma)\
 \sqrt{ 2}\ {\alpha N_c \over \pi}
\left({2a \over \pi}
\right)^{3/2} & \cr
 & \times \ {r^{-\gamma} \over x}\ {{\rm
exp} \left[
\left({\alpha N_c \over \pi} \chi (\gamma) - (\alpha_ p-1) \right)\ Y/2
\right] \over
2\ \left(\alpha_ p-1 \right)\
-\ {\alpha N_c \over \pi} \chi (\gamma)}
 & (18) \cr  &  \times
\ { {\rm ln}\left( {r \over x}\right)
\ {\rm exp}\left[-a {\rm ln}^2 \left({r
\over x} \right)\right]}
 \over 2 +
2 a {\rm ln} \left({r
\over x} \right)-\ \gamma
 &  \cr} $$
with
$$
{\cal H}(\gamma) = \int_0^1 \ {\rm d} y\ _2F_1(y^2)\left(1+y^{\gamma-2}\right)
.$$
To obtain
${ {\rm d}\hat \sigma \over {\rm d} x
{\rm d} r^2},$
the inverse Mellin transform of (18) has to be
determined. In performing it, one has to take into account the
poles in the $\gamma -$variable present in the denominators of
expression (18). In fact the one which is
less but closest to the $2 \ \pm i\infty$ line in the complex $\gamma$-plane
will dominate.
One obtains:
$${\alpha N_c \over \pi}\ \chi
\left(\gamma^{\ast}\right) =
2 (\alpha_p \ - \ 1 ) \equiv
\ {2\ \alpha N_c \over \pi}\ \chi(1),
$$
Denoting for convenience the solution $\gamma^{\ast} = 2\ - \gamma_M,$
and using the explicit form of the kernel $\chi(\gamma)$ leads to
equation (4) for $\gamma_M.$ Notice that the other denominator in (18)
corresponds to a pole outside the integration contour (at $\gamma > 2).$
In the same way, among the different solutions
of Eq.(4)  we justify the physical choice made in Ref.{4}, namely the
choice of the pole nearest and below $\gamma = 2.$

Using the relation:
$$
 r^{-\gamma}
\ { {\rm ln}\left( {r \over x}\right)
\ {\rm e}^{-a {\rm ln}^2 \left({r
\over x} \right)}
\ \over \gamma^{\ast}-\gamma }
 \equiv
\int_0^r x_{10}^{-\gamma-1}\ {\rm d}x_{10}
{\left(x_{10} \over r\right)}^{\gamma^{\ast}-
2 a {\rm ln} \left({r
\over x} \right)}\ \times
$$
$$ \  \ \  \   \times  \  {\rm ln}\left( {r^2 \over xx_{10}}\right)
\ {\rm e}^
{-a {\rm ln}^2 \left({r^2
\over xx_{10}} \right)},
\eqno (19)
$$
valid for ${r\over x} \gg 1,$
we obtain
formula  (1) as the inverse Mellin transform reciprocal to  (19).
The constant ${\cal C}$ in (1) is given by
$$
{\cal C} = {\sqrt 2\ {\cal H}(\gamma^{\ast})\ \left(2-\gamma{\ast}\right)
^{-1}
\over \psi^{\prime}(1-\gamma^{\ast}/2) - \psi^{\prime}(\gamma^{\ast}/2)}.
\eqno (20)
$$
It is important to notice that the integration range $0 \to r$
in (19) is required by the physical
constraint $x_{10} \ll r$ on the cross-section.\footnote{$^1$}
{The
approximation
$
a\ {\rm ln} \left({r
\over x} \right)\ll 1
$ has been used in the derivation of (1) from (18) and (20). A more
complete treatment is possible but is not relevant for the
discussion which always assumes the approximation scheme (5).}
This constraint is instrumental in the determination
of the appropriate cut-off $
{\rm ln}^2 \left({r^2
\over xx_{10}} \right)\ < \ a^{-1}.$

${\bf 3.}$ The most interesting feature of the cross-section expressions
(1)-(3) is the factor $
r^{\gamma_M}, $
responsible for the non-trivial power-law behaviour at rather large $ r
$ (at
least for
large enough incident energy). This factor is a direct consequence of the fact$
^{\lbrack 4\rbrack} $ that
the two-dipole density $ n^{(2)} $ in an onium state is not a mere direct
product of
two single dipole
densities $ n^{(1)}\otimes n^{(1)}. $ In other words, it is a consequence of
the scale-invariant {\it correlations}
between colour dipoles located at different positions in the transverse space
of the
onium. These correlations are -- in turn\nobreak\ -- to be considered as
consequences of the
(self-similar) cascading nature of dipole emission, as explicited in Mueller's
approach$^{[2-4]}.$

Since $ \gamma_M > 0, $ the power-law tail in (1)-(3) is not integrable
and thus the physical
distribution becomes  integrable only because of the cut-off $ r
<  r_{ {\rm max}}, $ see
(7), (22). Consequently, the Fourier transform of $ { {\rm d}\hat \sigma \over
{\rm d}^2r} $ reveals a power-law behaviour at
small transverse-momentum distance when it is greater than $ r^{-1}_{ {\rm
max}}. $
This feature gives interesting possibilities of observing
this component of soft gluon emission either in the form of
a power-law "spike" at very small transverse momenta
of the produced hadrons or as the power-law "singularity"
in HBT correlations between identical hadrons\footnote{$^2$}{
For incoherent emission, the HBT correlations
between momenta of identical particles are approximately
given by the square of the Fourier-transformed source density in space$^{[9]}.$
Therefore the power-law of the space density implies
the power-law singularity at small momentum difference$^{[10]}.$}.
One may even hope that the value of $\gamma_M$ can then be measured and
confronted with theory.
However, the feasibility of this program depends -
at least to some extent - on the mechanism of the hadronization
of the colour dipoles and therefore its discussion goes beyond the scope of the
present paper.
In particular a thorough discussion of the r\^ ole of the
dipole size $x$ is needed. Clearly further work and
other
physical tests of the predicted power law tail (before the cut-off) is
deserved,
especially in the experimental context of deep inelastic scattering at HERA,
where
conditions approaching onium-onium scattering can be realized.

Finally,  it is not excluded that effects
similar
to the ones computed in the theoretical framework of onium-onium scattering,
could be
present in hadronic collisions at high-energy. Obviously these
processes are
expected to be dynamically dominated by (yet uncalculable) non-perturbative
contributions.
However, if QCD Pomeron exchange (BFKL) is responsible for even a small part
of the
total hadron-hadron cross-section, the soft gluon emission of the type
described here
can be
the only contribution to the production process extending up to rather large
transverse
distances. If this is the case, it may
be possible to observe small but
clear effects
related to the power-law behaviour computed from perturbative QCD.
In this context, it is worth mentioning that  the existence of a perturbative
QCD tail of the
hadronic wave-function is advocated in some other cases, such
the proton form-factor at high $Q^2$ or the colour transparency effect
in reactions involving nuclei$^{[11]}.$
 However, the
limitations due to confinement forces must be better understood before definite
predictions can
be formulated.

In conclusion, we have computed the emission of colour dipoles
induced by QCD pomeron exchange in onium-onium scattering.
The resulting emission increases as a power of the ratio of the
center-of-mass energy over the light-cone momentum
of the softest gluon of the emitted dipole.
It shows the interesting feature of a power-law tail in the transverse distance
with respect to the collision axis, reflecting the conformal invariance of
the underlying theory.
\bigskip\bigskip
\noindent {\bf Acknowledgements}

It is a pleasure to thank A. Kaidalov, G. Korchemsky, Al. Mueller, H. Navelet
and S. Wallon
for inspiring discussions.

This work was partly supported by the KBN grant \# 2 PO3B 08308.
\vfill\eject

\centerline{{\bf References}}

\vglue 1truecm

\item{$\lbrack$1$\rbrack$}BFKL; Ya.Ya. Balitsky and L.N. Lipatov, {\it Sov. J.
Nucl. Phys.} {\bf 28} (1978)
822; E.A. Kuraev, L.N. Lipatov and V.S. Fadin, {\it Sov. Phys. JETP} {\bf 45}
(1997) 199; L.N.
Lipatov, {\it Sov. Phys. JETP} {\bf 63} (1986) 904.

\item{$\lbrack$2$\rbrack$}A.H. Mueller, {\it Nucl. Phys.} {\bf B415} (1994)
373.

\item{$\lbrack$3$\rbrack$}A.H. Mueller and B.Patel, {\it Nucl. Phys.} {\bf
B425} (1994) 471.

\item{$\lbrack$4$\rbrack$}A.H. Mueller, {\it Nucl. Phys.} {\bf B437} (1995)
107.

\item{$\lbrack$5$\rbrack$}A somewhat related approach has been proposed by N.N.
Nikolaev and B.G. Zakharov, {\it
JETP} {\bf 78} (1994) 598 and references therein.

\item{$\lbrack$6$\rbrack$}See Eq.(A.1) in Ref.$\lbrack$4$\rbrack$. We have
considered the $ Y $-differential version of this
equation, which is easily shown to be identical to the integrated version
of Ref.(4).
This
equivalence can be traced back to the equation defining the generating
function of
multi-dipole distributions.

\item{$\lbrack$7$\rbrack$}For instance: I.S. Gradsteyn and I.H. Rizhik,
Academic Press, $ 5^{ {\rm th}} $ edition,
Alan Jeffrey ed. (1994).

\item{$\lbrack$8$\rbrack$}
L.N. Lipatov {\it Phys. Lett.} {\bf B251} (1980) 413,
{\bf B309} (1993) 394.
L. Faddeev and G. P. Korchemsky,
{\it Phys. lett.} {\bf B342} (1994) 311.

\item{$\lbrack$9$\rbrack$}
Hanbury-Brown and Twiss effect: for a review see D.H. Boal et al.,
{\it Rev. Mod. Phys.} {\bf 62} (1990) 553.

\item{$\lbrack$10$\rbrack$}
A. Bialas, {\it Acta Phys. Pol.} {\bf B23} (192) 561.

\item{$\lbrack$11$\rbrack$}
See for instance,
{\it Basics of Perturbative QCD}, Yu.L. Dokshitzer, V.A.
Khoze, A.H. Mueller and S.I. Troyan, (J. Tran Than Van ed., Editions Fronti\`
eres) 1991, and references therein.

\vfill\eject

\centerline{{\bf Figure Caption}}

\vglue 1truecm

\item{Fig.1 :}Transverse-Plane Geometry of Dipole-Emission.

\item{\nobreak\ \nobreak\ \nobreak\ \nobreak\ \nobreak\ \nobreak\ \nobreak\
}The dashed area around each dipole is of a typical size of the order of
the associated dipole length. They are assumed to be much smaller than the
ranges in
$ \vec r, $ $ \vec b, $ $ \vec b^{\prime} , $
$ \vec b^{\prime} -\vec b.$

\par\vfill\supereject\end